\documentclass[prd]{revtex4-1}
\usepackage{amsmath}  
\usepackage{amsfonts} 
\usepackage{graphicx} 
\usepackage{amsmath,amssymb,epsfig}
\usepackage{graphicx}
\usepackage[vcentermath,stdtext]{youngtab}
\usepackage{epstopdf}

\begin{document}

\title{Role of pentaquark components in $\phi$ meson production proton-antiproton annihilation reactions}

\author{Sorakrai Srisuphaphon$^1$, A. Kaewsnod$^2$, A. Limphirat$^{2,3}$,
K. Khosonthongkee$^{2,3}$, Y. Yan$^{2,3}$\footnote{Corresponding author: yupeng@sut.ac.th}}
\affiliation{$^1$Department of Physics, Faculty of Science, Burapha university, Chonburi, 20131, Thailand, $^2$ School of Physics, Suranaree
University of Technology, 111 University Avenue, Nakhon Ratchasima
30000, Thailand
\\$^3$
Thailand Center of Excellence in Physics, Ministry
of Education, Bangkok, Thailand} 

\date{\today}

\begin{abstract}
The pentaquark component $uuds\bar s$ is included in the proton
wave functions to study $\phi $ meson production proton-antiproton
annihilation reactions. With all possible configurations of the $uuds$ subsystem proposed
for describing the strangeness spin and magnetic moment of the proton,
we estimate the branching ratios of the
annihilation reactions at rest $p\bar{p} \rightarrow \phi X$
($X=\pi^0,\,\eta,\,\rho^0,\,\omega$) from atomic $p\bar{p}$ $S$- and
$P$-wave states by using effective quark line diagrams incorporating the
$^3P_0$ model. The best agreement of theoretical prediction with the experimental data is found
when the pentaquark  configuration of the proton wave function takes the flavor-spin symmetry $[4]_{FS}[22]_F[22]_S $.
\end{abstract}


\maketitle 

\newpage

\section{Introduction}\label{sec:1}

 The apparent OZI rule violation in $\phi$ production nucleon-antinucleon annihilation reactions suggests the existence of strange quarks in the nucleon \cite{Amsler:1991hs,Amsler:1997up}.
There are also other experimental results indicating the strangeness content in the nucleon, for example, the strange quark-antiquark contributions to the electric and magnetic strange form factors of the nucleon in the parity violation experiments of electron scattering from nucleon \cite{Armstrong:2012bi}. The strangeness magnetic moment $\mu_s$ obtained by extrapolating the magnetic strange form factors $G_M^s(Q^2)$ at momentum transfer $Q^2=0$ suggest a positive value for $\mu_s$ \cite{Diehl:2007uc}, but most theoretical calculations in the $3q$ picture of the proton derive a negative value for this observable as reviewed in Ref \cite{Beck:2001yx,Green:2015wqa}.

 There is an interesting work, where the strangeness magnetic moment of the proton is studied by including the pentaquark component $uuds\bar{s}$, in addition to $3q$ component ($uud$), into the proton wave function \cite{An:2005cj}. It is shown in Ref. \cite{An:2005cj} that almost all $5q$ configurations give the strangeness spin contribution $\sigma_s $ a negative value, consistent with the experimental and theoretical indications of the spin structure of the proton \cite{Filippone:2001ux}. However, only the $5q$ configurations where the $\bar{s}$ is in the $P$-state with $uuds$ subsystem is in their ground state result in negative values for $\mu_s$, while positive values for $\mu_s$ are found in the $5q$ configurations where $\bar{s}$ is in ground state but the subsystem $uuds$ is in the $P$-state. The proposed pentaquark picture has been applied to other works such as the estimation of admixture $uuds\bar{s}$ component in the nucleon wave function with the strangeness form factor of the proton \cite{Riska:2005bh,Kiswandhi:2011ce,An:2013daa}, the study of the amplitudes for the electromagnetic transition $\gamma^* N \rightarrow N^*(1535)$ \cite{An:2008xk} and the electromagnetic decay of the $N(1440)$ resonance \cite{JuliaDiaz:2006av,Li:2006nm}.
	
 In the present work we study $\phi$ meson production proton-antiproton annihilation reactions $p\bar{p}\rightarrow \phi X$ ($X=\pi^0,\eta,\rho^0,\omega$), considering all possible $uuds\bar{s}$ configurations for the proton wave function in addition to the $3q$ component. In our previous work \cite{Srisuphaphon:2011xn}, the proton-antiproton annihilation reactions have been studied with the $5q$ components in three models, namely, the $uud$ cluster with a $s\bar{s}$ sea quark component, kaon-hyperon clusters based on the chiral quark model, and the pentaquark picture $uuds\bar s$ where $\bar{s}$ being in the $P$-state while $uuds$ subsystem in their ground state. In the case of the pentaquark picture, two configurations of $uuds\bar{s}$, the mixed flavor-spin symmetries $[31]_{FS}[211]_F[22]_S$ and $[31]_{FS}[31]_F[22]_S$ were considered since these two configurations result in negative values for $\sigma_s  $ and $\mu_s$. The theoretical branching ratios in comparison with experimental data for these two configurations were discussed and listed in Table III in Ref \cite{Srisuphaphon:2011xn}.

 There are 15 possible $5q$ configurations, where the $\bar{s}$ is in the ground state but the subsystem $uuds$ is in the $P$-state, resulting in negative values for the $\sigma_s$. It is shown in Ref \cite{An:2005cj} that the configuration with $[4]_{FS}[22]_F[22]_S$ flavor-spin symmetry is likely to have the lowest energy. In this work, we study $\phi$ meson production proton-antiproton annihilation reactions $p\bar{p}\rightarrow \phi X$ ($X=\pi^0,\eta,\rho^0,\omega$), focusing on these 15 configurations. The paper is organized as follows. Proton wave functions with various $5q$ component are constructed in Section 2. In Section 3 we evaluate the branching ratios for the reactions $p\bar{p}\rightarrow \phi X$. Finally a summary and conclusions are given in Section 4.

\section{Proton wave functions with pentaquark components}\label{sec:2}
The proton wave function with the addition of $uuds \bar{s}$ components to the $uud$ quark component may be written generally in the form \cite{Ellis:1994ww}
\begin{equation}
|p\rangle = A|uud\rangle+B|uuds\bar{s}\rangle,
\end{equation}
where $A$ and $B$ are the amplitude factors for the $uud$ and $uuds \bar{s}$  components in the proton, respectively.
The 5$q$ states may constructed by coupling the $uuds$ wave function with the $\bar{s}$ one.

In the language of group theory, the permutation symmetry of the
four-quark configuration is characterized by the $S_4$ Young
tabloids $[4]$, $[31]$, $[22]$, $[211]$ and $[1111]$. That the pentaquark wave function
should be a color
singlet demands that the color part of the pentaquark wave function
must be a $[222]_1$ singlet. The color
state of the antiquark in pentaquarks is a $[11]$ antitriplet,
thus the color wave function of the four-quark configuration
must be a $[211]_3$ triplet,
\begin{eqnarray}
\chi^c_{[211]_\lambda}(q^4)=\young(12,3,4)\;\;\;\;\chi^c_{[211]_\rho}(q^4)=\young(13,2,4)\;\;\;\;\chi^c_{[211]_\eta}(q^4)=\young(14,2,3).
\end{eqnarray}
The $q^4$ color wave functions can be derived by
applying the $\lambda$-type, $\rho$-type and $\eta$-type projection
operators of the irreducible representation IR[211] of the permutation group $S_4$ in Yamanouchi basis, onto
single particle color states. The details can be found in Ref.\cite {Yan:2012zzb}.
For the product state $RRGB$, for example, we have,
\begin{eqnarray}
&& P_{[211]_\lambda}(RRGB) \Longrightarrow \chi^c_{[211]_\lambda}(R), \nonumber \\
&& P_{[211]_\rho}(RGRB) \Longrightarrow \chi^c_{[211]_{\rho}}(R), \nonumber \\
&& P_{[211]_\eta}(RGBR) \Longrightarrow \chi^c_{[211]_\eta}(R),
\end{eqnarray}
with
\begin{eqnarray}\label{C1-222}
\chi^c_{[211]_\lambda}(R) &=& \frac{1}{\sqrt{16}}
(2|RRGB\rangle-2|RRBG\rangle \nonumber \\
&& -|GRRB\rangle
-|RGRB\rangle -|BRGR\rangle
-|RBGR\rangle \nonumber \\
&& +|BRRG\rangle+|GRBR\rangle
+|RBRG\rangle +|RGBR\rangle), \nonumber
\end{eqnarray}
\begin{eqnarray}\label{C3-222}
\chi^c_{[211]_{\rho}}(R) & =& \frac{1}{\sqrt{48}}
(3|RGRB\rangle-3|GRRB\rangle \nonumber \\
&& +3|BRRG\rangle
-3|RBRG\rangle +2|GBRR\rangle -2|BGRR\rangle \nonumber \\
&& -|BRGR\rangle+|RBGR\rangle
+|GRBR\rangle-|RGBR\rangle), \nonumber
\end{eqnarray}
\begin{eqnarray}\label{C3-222}
\chi^c_{[211]_\eta}(R) &=& \frac{1}{\sqrt{6}}
(|BRGR\rangle+|RGBR\rangle+|GBRR\rangle  \nonumber \\
&& -|RBGR\rangle-|GRBR\rangle-|BGRR\rangle).
\end{eqnarray}
Thus, the corresponding singlet color wave function of the pentaquark at color symmetry pattern $j=\lambda,\rho,\eta$ is given by
\begin{eqnarray}\label{color-uuds}
\chi^C_{[222]_j}=\frac{1}{\sqrt{3}}\left[\chi^c_{[211]_j}(R)\,\bar{R}+\chi^c_{[211]_j}(G)\,\bar{G}+\chi^c_{[211]_j}(B)\,\bar{B}\right].
\end{eqnarray}

\begin{table}\label{osfConfigurations}
\caption{Spatial-spin-flavor configurations of $q^4$ clusters}
\begin{center}
\vspace*{.3cm}
\begin{tabular}{c c}
\hline
\multicolumn{2}{c}{$[31]_{OSF}$}\\
   \hline\multicolumn{1}{l}{$[4]_{O}$}& \multicolumn{1}{l}{$[31]_{SF}$}\\
         \multicolumn{1}{l}{$[1111]_{O}$}& \multicolumn{1}{l}{$[211]_{SF}$}\\
         \multicolumn{1}{l}{$[22]_{O}$}& \multicolumn{1}{l}{$[31]_{SF},[211]_{SF}$}\\
         \multicolumn{1}{l}{$[211]_{O}$}& \multicolumn{1}{l}{$[31]_{SF},[211]_{SF},[22]_{SF}$}\\
         \multicolumn{1}{l}{$[31]_{O}$}& \multicolumn{1}{l}{$[4]_{SF},[31]_{SF},[211]_{SF},[22]_{SF}$}\\
   \hline
\end{tabular}
\end{center}
\end{table}

The total wave function of the four quark configuration is
antisymmetric implies that the spatial-spin-flavor part must be a [31]
state. The total wave function of the $q^4$ configuration may be written in
the general form
\begin{eqnarray}
\psi_{[1111]}=\sum_{i,j=\lambda,\rho,\eta}a_{ij}\; \chi^c_{[211]_i}
\chi^{osf}_{[31]_j}.
\end{eqnarray}
The coefficients $a_{ij}$ can be determined easily by applying the IR[31] and IR[211] of the permutation group $S_4$ in Yamanouchi basis onto the equation. The total wave function of the $q^4$ subsystem takes the form,
\begin{eqnarray}
\psi_{[1111]} = \frac{1}{\sqrt{3}}
\left( \chi^{c}_{[211]_\lambda} \chi^{osf}_{[31]_\rho} -
\chi^{c}_{[211]_\rho} \chi^{osf}_{[31]_\lambda} +
\chi^{c}_{[211]_\eta} \chi^{osf}_{[31]_\eta} \right).
\end{eqnarray}
The spatial-spin-flavor and spin-flavor states of the $q^4$ cluster in the above equation can be written
in the general forms,
\begin{eqnarray}\label{eqn::osf}
\chi^{osf}_{[31]} &=& \sum_{i,j}a_{ij}\chi^{o}_{[X]_{i}}\chi^{sf}_{[Y]_{j}},
\\
\chi^{sf}_{[Z]} &=& \sum_{i,j}a_{ij}\chi^{f}_{[X]_{i}}\chi^{s}_{[Y]_{j}},
\end{eqnarray}
where $\chi^{o}_i$, $\chi^{f}_i$ and $\chi^{s}_i$ are respectively the $q^4$ spatial, flavor and spin wave functions of symmetry (S), antisymmetry (A), $\lambda$-type, $\rho$-type and $\eta$-type.
The possible spatial-spin-flavor and spin-flavor configurations and explicit forms of the wave functions
are determined by applying the $S_4$ representations in Yamanouchi basis.
Listed in Table \ref{osfConfigurations} and Table \ref{sfConfigurations} are respectively
the possible spatial-spin-flavor and spin-flavor configurations.

\begin{table}\label{sfConfigurations}
\caption{Spin-flavor configurations of $q^4$ clusters}
\begin{center}
\vspace*{.3cm}
\begin{tabular}{c c c c c}
\hline\multicolumn{3}{c}{$[4]_{FS}$}\\&\\
          \multicolumn{1}{l}{$[22]_{F}[22]_{S}$}&
          \multicolumn{1}{l}{$[31]_{F}[31]_{S}$}&
          \multicolumn{1}{l}{$[4]_{F}[4]_{S}$}\\
\hline\multicolumn{3}{c}{$[31]_{FS}$}\\&\\
          \multicolumn{1}{l}{$[31]_{F}[22]_{S}$}&
          \multicolumn{1}{l}{$[31]_{F}[31]_{S}$}&
          \multicolumn{1}{l}{$[31]_{F}[4]_{S}$}&
          \multicolumn{1}{l}{$[211]_{F}[22]_{S}$}\\&\\
          \multicolumn{1}{l}{$[211]_{F}[31]_{S}$}&
          \multicolumn{1}{l}{$[22]_{F}[31]_{S}$}&
          \multicolumn{1}{l}{$[4]_{F}[31]_{S}$}\\
   \hline\multicolumn{3}{c}{$[22]_{FS}$}\\&\\
          \multicolumn{1}{l}{$[22]_{F}[22]_{S}$}&
          \multicolumn{1}{l}{$[22]_{F}[4]_{S}$}&
          \multicolumn{1}{l}{$[4]_{F}[22]_{S}$}&
          \multicolumn{1}{l}{$[211]_{F}[31]_{S}$}\\&\\
          \multicolumn{1}{l}{$[31]_{F}[31]_{S}$}\\
   \hline\multicolumn{3}{c}{$[211]_{FS}$}\\&\\
          \multicolumn{1}{l}{$[211]_{F}[22]_{S}$}&
          \multicolumn{1}{l}{$[211]_{F}[31]_{S}$}&
          \multicolumn{1}{l}{$[211]_{F}[4]_{S}$}&
          \multicolumn{1}{l}{$[22]_{F}[31]_{S}$}\\&\\
          \multicolumn{1}{l}{$[31]_{F}[22]_{S}$}&
          \multicolumn{1}{l}{$[31]_{F}[31]_{S}$}\\
   \hline
\end{tabular}
\end{center}
\end{table}
According to the requirement of positive parity for the proton wave function, if the $uuds$ subsystem is in the ground state then $\bar{s}$ has to be in the P-state.
For the $uuds$ subsystem in the ground state, the spatial part of the subsystem takes the $[4]_O$ symmetry and hence the spin-flavor part must take the $[31]_{FS}$ symmetry as shown in Table \ref{osfConfigurations} in order to form an antisymmetric spatial-color-spin-flavor $uuds$ part of the pentaquark wave function.  If the spin symmetry of the $uuds$ subsystem is described by $[22]_S$ corresponding to spin 0, the flavor symmetry representations $[31]_F$ and $[211]_F$ may combine with the spin symmetry state $[22]_S $ to form the mixed symmetry spin-flavor states $[31]_{FS}$ as shown in Table \ref{sfConfigurations} here and in Ref. \cite{An:2005cj}. In this case, the 5q component may be written in the general form
\begin{eqnarray}\label{5q-csf0}
|uuds\bar{s }\rangle=[|\frac{1}{2},m_{\bar{s}}\rangle \otimes |1,\mu\rangle]_{\frac{1}{2},m_{5q}}
\frac{1}{\sqrt{3}}\big(\chi^C_{[222]_{\lambda}}(\bar{s}\chi^{FS}_{[31]_\rho})
~-\chi^C_{[222]_{\rho}}(\bar{s}\chi^{FS}_{[31]_\lambda})~+\chi^C_{[222]_{\eta}}(\bar{s}\chi^{FS}_{[31]_\eta})\big),
\end{eqnarray}
where $(\frac{1}{2},m_{5q})$ denotes the spin of the 5q component. The spin state of $\bar s$ and the angular momentum $\ell=1$ are denoted by $|\frac{1}{2},m_{\bar s}\rangle$ and $|1,\mu\rangle$, respectively. The function $\chi^{FS}_{[31]_{(\lambda, \rho, \eta)}}$ represents the coupled spin-flavor part with the mixed symmetry $[31]_{FS}$. The $5q$ component with the configurations $[31]_{FS}[211]_{F}[22]_{S}$ and $[31]_{FS}[31]_{F}[22]_{S} $ results in negative values for $\sigma_s $ and $\mu_s $ \cite{An:2005cj}.

For the $P$-state $uuds$ subsystem, in the language of group theory, the orbital angular momentum $\ell=1$ means that the spatial wave function of the subsystem has the mixed symmetry $[31]_O$. Therefore, the possible spin-flavor configurations are $ [4]_{FS},\, [31]_{FS},\, [211]_{FS}$ and $ [22]_{FS}$, as shown in Table \ref{osfConfigurations}, which couple with the $[31]_{O}$ spatial state to form the $[31]_{OSF}$ spatial-flavor-spin components of the $uuds$ subsystem. There are three possible spin symmetries of the $uuds$ subsystem: $[22]_S$, $[31]_S$ and $[4]_S $ representations as shown in Table \ref{sfConfigurations}, corresponding to spin $S=0,\,1,\,2$, respectively. For this case, the full wave functions of the 5q component will be presented in Section 3.

\section{The $N \bar N$ transition amplitude and branching ratios}\label{sec:3}
In this work we study the annihilation reactions $N\bar N \to X \phi $
$(X=\pi^0 , \eta ,\rho^0, \omega )$ with the effective quark line diagram for the shake-out of a $\phi$ meson from the $5q$ component, as described in Fig.1 \cite{Ellis:1994ww}.

            \begin{figure}[h!]
            \begin{center}
             \includegraphics[width=7cm]{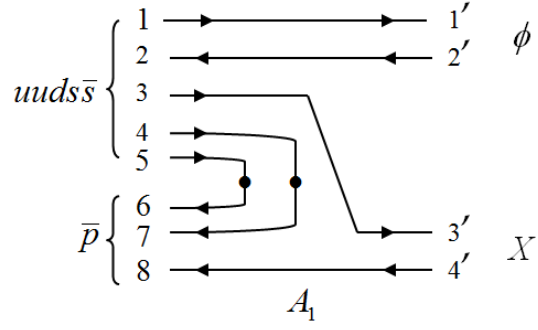}
            \end{center}
            \caption{The quark line diagram corresponding to the production of two meson final states in $p\bar p$ annihilation \cite{Ellis:1994ww,Gutsche:1997gy}. The dots refer to the effective vertex of the $^3P_0$ for $q\bar q$ pairs are destroyed with the quantum numbers of the vacuum : $^3P_0$, isospin $I=0$ and color singlet \cite{LeYaouanc:1988fx}.}
            \end{figure}

According to the quark diagram, the transition amplitudes from the $5q$ component $|uuds\bar{s }\rangle$ and the antiproton $|\bar{u} \bar{u} \bar{d}\rangle$ wave function in the momentum space representation is given by
\begin{eqnarray} T_{A_1} =2AB \int d^3q_1
..d^3q_8 d^3q_{1'}..d^3q_{4'}\langle \phi
X|\vec{q}_{1'}..\vec{q}_{4'} \rangle
\langle\vec{q}_{1'}..\vec{q}_{4'}|
\mathcal{O}_{A_1}|\vec{q}_{1}..\vec{q}_{8}\rangle  \langle
\vec{q}_{1}..\vec{q}_{8}|(uuds\bar{s})\otimes(\bar{u} \bar{u}
\bar{d})\rangle.
\end{eqnarray}
The effective operators $\mathcal{O}_{A_I}$, is corresponding to the quark line diagram, take the form
\begin{equation}
\mathcal{O}_{A_1} = \lambda_{A_1}
\delta^{(3)}(\vec{q}_1-\vec{q}_{1'})
\delta^{(3)}(\vec{q}_2-\vec{q}_{2'})\delta^{(3)}(\vec{q}_3-\vec{q}_{3'})\delta^{(3)}(\vec{q}_8-\vec{q}_{4'})V^{56}V^{47},
\end{equation}
where $\lambda_{A_1}$ is a parameter describe the effective strength of the transition topology which can be fitted by a experimental data. The $^3P_0$ quark-antiquark vertex is defined as
\begin{equation}
V^{ij}=\sum_\mu\sigma^{ij}_{-\mu}Y_{1\mu}(\vec{q}_i-\vec{q}_{j})\delta^{(3)}(\vec{q}_i+\vec{q}_{j})(-1)^{1+\mu}1^{ij}_F1^{ij}_C~,
\end{equation}
where $\sigma^{ij}_{-\mu}$ is the spin operator for destroying $q_i \bar{q}_j$ pairs with spin 1 while  $Y_{1\mu}(\vec{q})$ is the spherical harmonics in momentum space \cite{Dover:1992vj}. The unit
operators in flavor and color spaces are denoted by $1^{ij}_{F}$ and $1^{ij}_{C}$, respectively. Nevertheless, the 5$q$
component had been treated as a small perturbative admixture in the
proton ($B^2<<1$), the transition amplitude with the term
$ \langle \vec{q}_{1}..\vec{q}_{8}|(uuds\bar{s})\otimes(\bar{u}
\bar{u} \bar{d}\bar{s}s)\rangle$ corresponding to the rearrangement process
\cite{Ellis:1994ww} can be ignored.

For the mesons $ M $ ($ \phi$ and
$ X $) and $\bar p$ ($\bar{q}^3$), the corresponding wave function can be expressed in terms of the quark
momenta as
\begin{equation}
\langle \vec{q}_{i'}
\vec{q}_{j'} | M\rangle\equiv\varphi_M(\vec{q}_{i'},\vec{q}_{j'})\chi_{M}(q\bar{q})=N_M{\rm
exp}\left\{-\frac{R^2_M}{8} \Big(\vec{q}_{i'} -\vec{q}_{j'}\Big)^2
\right\}\chi_M(q\bar{q}),
\end{equation}
\begin{equation}
\langle \vec{q}_6 \vec{q}_7 \vec{q}_8|\bar{u} \bar{u}
\bar{d}\rangle\equiv\varphi_{\bar p}(\vec{q}_6, \vec{q}_7, \vec{q}_8)\chi_{\bar p}(\bar{q}^3)=
N_B{\rm exp}\left\{-\frac{R^2_B}{4}
\Big[(\vec{q}_7-\vec{q}_8)^2
+\frac{(\vec{q}_7+\vec{q}_8-2\vec{q}_6)^2}{3}\Big]  \right\}
\chi_{\bar p}(\bar{q}^3),
\end{equation}
respectively, where $N_{M} = (R_{M}^2/\pi)^{3/4}$, $N_B=(3R^2_B/\pi)^{3/2}$ with $R_{(M,B)}$ being the meson (baryon) radial parameter. Here, $\chi_M(q\bar{q})$ and $\chi_{B}(\bar{q}^3)$ denote the spin-flavor-color wave function: $[S]\otimes[F]\otimes[C]$. Note that the internal spatial wave functions are approximated as the
harmonic oscillator forms.


In case of $uuds$ quarks in their ground state with the spatial state symmetry $[4]_O$, the full $5q$ component wave function is given by
\begin{eqnarray} \label{WF0}
\langle \vec{q}_1...\vec{q}_5 |uuds\bar{s} \rangle
=\varphi_{uuds\bar{s} }(\vec{q}_1,..,\vec{q}_5)~Y_{1\mu}\biggl(\frac{\vec{q}_2+\vec{q}_3+\vec{q}_4
+\vec{q}_5-4\vec{q}_1}{\sqrt{20}}\biggr)~\psi_{uuds\bar{s }},
\end{eqnarray}
with
\begin{eqnarray}\label{rad}
\varphi_{uuds\bar{s} }(\vec{q}_1,..,\vec{q}_5)
=
N_{uuds\bar{s}} \,
{\rm exp}\biggl\{-\frac{R^2_{uuds\bar{s }}}{4}\Big[(\vec{q}_2-\vec{q}_3)^2
+\frac{(\vec{q}_2+\vec{q}_3-2\vec{q}_4)^2}{3}~~~~~~~
\nonumber\\
~~~~~+\frac{(\vec{q}_2+\vec{q}_3+\vec{q}_4-3\vec{q}_5)^2}{6}+
\frac{(\vec{q}_2+\vec{q}_3+\vec{q}_4+\vec{q}_5-4\vec{q}_1)^2}{10}\Big]\biggr\},
\end{eqnarray}
where $\psi_{uuds\bar{s }}$ is the spin-flavor-color wave function as defined in Eq.(\ref{5q-csf0}).

The $5q$ wave function for the $uuds$ subsystem with the orbital angular momentum $\ell=1$ can be constructed systematically in the group theory approach mentioned in Section 2. For instance, for the simplest case where the spin-flavor part of the $q^4$ subsystem takes the $[4]_{FS}$ symmetry, the wave function takes the form,
\begin{eqnarray} \label{WF}
\langle uuds\bar{s} |\vec{q}_1...\vec{q}_5 \rangle
=\varphi_{uuds\bar{s} }(\vec{q}_1,..,\vec{q}_5)\psi_{uuds\bar{s}},
\end{eqnarray}
where $\varphi_{uuds\bar{s} }(\vec{q}_1,\cdots,\vec{q}_5)$ has the same form as shown in Eq.(\ref{rad}). Here, $\psi_{uuds\bar{s }}$ representing the spatial-spin-flavor-color wave function of the $5q$ component is derived as
\begin{eqnarray}\label{5q-SC-WF}
\psi_{uuds\bar{s }}
= \sum_{J_{4q}}\left[ | \frac{1}{2},m_{\bar s}\rangle\otimes\left[\bar s\,\chi^{FS}_{[4]_{S_{4q},m_{4q}}}(uuds)\otimes\chi^{OC}_{\ell,\mu}\right]_{J_{4q},m_{4q}}\right]_{\frac{1}{2},m_{5q}}
\end{eqnarray}
where
\begin{eqnarray}\label{5q-OC-WF}
\chi^{OC}_{1,\mu} = \frac{1}{\sqrt{3}}
\left[\chi^C_{[222]_{\lambda}}Y_{1\mu}\biggl(\frac{\vec{q}_2-\vec{q}_3
}{\sqrt{2}}\biggr)-\chi^C_{[222]_{\rho}}Y_{1\mu}\biggl(\frac{\vec{q}_2+\vec{q}_3-2\vec{q}_4
}{\sqrt{6}}\biggr)+\chi^C_{[222]_{\eta}}Y_{1\mu}\biggl(\frac{\vec{q}_2+\vec{q}_3+\vec{q}_4-3\vec{q}_5
}{\sqrt{12}}\biggr) \right] \nonumber \\
\end{eqnarray}
for $\ell=1$, and $J_{4q}=\ell \oplus S_{4q}$ is the total angular momentum for the $uuds$ subsystem.

There are three configurations for the $q^4$ spin-flavor symmetry $[4]_{FS}$, that is, $[22]_{F}[22]_{S}$, $[31]_{F}[31]_{S}$ and $[4]_{F}[4]_{S}$ as shown in Table \ref{sfConfigurations}, and hence three $q^4$ spin-flavor wave functions $\chi^{FS}_{[4]}$ as follows:
\begin{eqnarray}
\chi^{FS}_{[4]_{S_{4q}=0}}&=& \sum_{i,j}a_{ij}\chi^{F}_{[22]_{i}}\chi^{S}_{[22]_{j}}, \nonumber \\
\chi^{FS}_{[4]_{S_{4q}=1}}&=& \sum_{i,j}a_{ij}\chi^{F}_{[31]_{i}}\chi^{S}_{[31]_{j}}, \nonumber \\
\chi^{FS}_{[4]_{S_{4q}=2}}&=& \sum_{i,j}a_{ij}\chi^{F}_{[4]_{i}}\chi^{S}_{[4]_{j}}.
\end{eqnarray}
It is an easy task to determine the coefficients by applying the $S_4$ representations in Yamanouchi basis onto the above equations. The explicit forms of the spin and flavor wave functions can be work out in the approach of projection operators as shown in Section 2 for the color wave functions.

  In the present work we consider the $\phi$ meson production $p \bar p$ annihilation reactions with the $5q$ component for the case of the subsystem $uuds$ in the $P$-state. 15 possible configurations of $uuds \bar{s}$ will be considered. In order to involve relative motion, we a choose plane wave basis for the relative motions of $p \bar p$ and $\phi X$ in the center of momentum system : $\delta^{(3)}(\vec{q}_1+\vec{q}_2+\vec{q}_3+\vec{q}_4+\vec{q}_5-\vec{k})\delta^{(3)}(\vec{k}+\vec{q}_6+\vec{q}_7+\vec{q}_8)$ and $\delta^{(3)}(\vec{q}-\vec{q}_{1'}-\vec{q}_{2'})\delta^{(3)}(\vec{q}+\vec{q}_{3'}+\vec{q}_{4'})$, respectively. In the low-momentum approximation as done in Ref \cite{Srisuphaphon:2011xn}, the leading order of the transition amplitude $T_{fi}^{SP(PS)}$ for the $S$ to $P$ ($L=0,$ $\ell_f=1$) and $P$ to $S$  ($L=1,$ $\ell_f=0$) transitions from the initial state $|i\rangle $ to final state $|f\rangle $ with the quark line diagram $A_1$ can be obtained as
\begin{eqnarray}\label{T-1}
T_{fi}^{SP(PS)}(\vec{q},\vec{k})=2AB \lambda_{A_1}N \pi^4 q^{l_f}k^{L} {\rm exp} \left\{ -Q^2_q q^2 -Q^2_k k^2\right\}\langle f | O_{A_1}|i \rangle,
\end{eqnarray}
where $N=N_\phi N_X N_{uuds\bar{s} } N_{\bar{p} }$ and
\begin{equation}\label{SFweight}
\langle f | O_{A_1}|i \rangle=\langle f | \sum_{\nu,\lambda} (-1)^{\nu+\lambda} \sigma^{56}_{-\nu}\sigma^{47}_{-\lambda}1^{56}_F1^{47}_F 1^{56}_C1^{47}_C \sum_{{m,n}}\Omega_{m,n}^{SP(PS)}f_{m,n}^{SP(PS)}(\nu,\mu,\lambda,L,M,l_f,m_f)|i \rangle,
\end{equation}
is the spin-color-flavor weight. The spin-angular momentum functions $f_{m,n}^{SP(PS)}(\nu,\mu,\lambda,L,M,l_f,m_f)$ are given by
\begin{eqnarray}\label{f-fn}
f_{1,1}^{SP}=f_{1,2}^{SP}=f_{1,3}^{SP}=(-1)^{\nu}\delta_{\lambda,-\nu}\delta_{\mu,m_f},~~~~~~~~~~~~
\nonumber \\f_{2,1}^{SP}=f_{2,2}^{SP}=f_{2,3}^{SP}=(-1)^{\mu}\delta_{\mu,-\nu}\delta_{\lambda,m_f},~~~~~~~~~~~~
\nonumber \\f_{3,1}^{SP}=f_{3,2}^{SP}=f_{3,3}^{SP}=(-1)^{\mu}\delta_{\mu,-\lambda}\delta_{\nu,m_f},~~~~~~~~~~~~
\nonumber \\f_{1,1}^{PS}=f_{1,2}^{PS}=f_{1,3}^{PS}=(-1)^{\lambda+\mu}\delta_{\lambda,-\nu}\delta_{\mu,-M},~~~~~~~~~~~~
\nonumber \\f_{2,1}^{PS}=f_{2,2}^{PS}=f_{2,3}^{PS}=(-1)^{\lambda+\mu}\delta_{\mu,-\nu}\delta_{\lambda,-M},~~~~~~~~~~~~
\nonumber \\f_{3,1}^{PS}=f_{3,2}^{PS}=f_{3,3}^{PS}=(-1)^{\nu+\mu}\delta_{\mu,-\lambda}\delta_{\nu,-M}.~~~~~~~~~~~~
\end{eqnarray}
The geometrical constants $Q^2_k, Q^2_q $ and $ \Omega_{m,n}^{SP(PS)} $ depending on the meson and baryon size parameters are given as followings:

\allowdisplaybreaks
\begin{eqnarray}\label{the constants1}
&& Q_k^2 = -\frac{R_M^4}{9 \left(3
   \left(R_B^2+R_{uuds\bar{s}}
   ^2\right)+2
   R_M^2\right)}+\frac{R_M^2}{
   18}+\frac{R_{uuds\bar{s}}^2
   }{15}
\nonumber \\
&&Q_p^2 = \frac{12 R_B^2 \left(R_M^2+6
   R_{uuds\bar{s}}^2\right)+R_{uuds\bar{s}}^2 \left(28
   R_M^2+15 R^2+25
   R_{uuds\bar{s}}^2\right)}{3
   2 \left(3 R_B^2+2 R_M^2+3
   R_{uuds\bar{s}}^2\right)},
\nonumber \\
&&\Omega_{1,1}^{SP}=-\frac{\sqrt{3} \left(\beta _2+4 \beta _3+4 \beta _4+3\right) \left(2
   Q_3-Q_4\right) \left(2 Q_3+Q_4\right)}{8 Q_2^3 Q_3^5 Q_4^5},
\nonumber \\
&&\Omega_{2,1}^{SP}=-\frac{\sqrt{3} \left(\beta _3+\beta _4+1\right) \left(2 Q_3-Q_4\right) \left(2
   Q_3+Q_4\right)}{2 Q_2^3 Q_3^5 Q_4^5},
\nonumber \\
&&\Omega_{3,1}^{SP}=-\frac{\sqrt{3} \beta _4 \left(4 Q_3^2+Q_4^2\right)}{2 Q_2^3 Q_3^5 Q_4^5},
\nonumber \\
&&\Omega_{1,2}^{SP}=-\frac{\sqrt{\frac{3}{2}} \left(\beta _2+\beta _3-2 \beta _4\right) \left(4
   Q_3^2-Q_4^2\right)}{4 Q_2^3 Q_3^5 Q_4^5},
\nonumber \\
&&\Omega_{2,2}^{SP}=\frac{\sqrt{\frac{3}{2}} \left(\beta _3+\beta _4+1\right) \left(2
   Q_3^2+Q_4^2\right)}{Q_2^3 Q_3^5 Q_4^5},
 \nonumber \\
&&\Omega_{3,2}^{SP}=\frac{\sqrt{\frac{3}{2}} \beta _4 \left(2 Q_3^2-Q_4^2\right)}{Q_2^3 Q_3^5 Q_4^5},
\nonumber \\
&&\Omega_{1,3}^{SP}=-\frac{3 \left(\beta _2-\beta _3\right) \left(4 Q_3^2-Q_4^2\right)}{4 \sqrt{2}
   Q_2^3 Q_3^5 Q_4^5},
\nonumber \\
&&\Omega_{2,3}^{SP}=-\frac{3 \left(\beta _3+\beta _4+1\right)}{2 \sqrt{2} Q_2^3 Q_3^5 Q_4^3},
\nonumber \\
&&\Omega_{3,3}^{SP}=\frac{3 \beta _4}{2 \sqrt{2} Q_2^3 Q_3^5 Q_4^3},
\nonumber \\
&&\Omega_{1,1}^{PS}=-\frac{\sqrt{3} \left(\alpha _2+4 \alpha _3+4 \alpha _4-3\right) \left(4
   Q_3^2-Q_4^2\right)}{8 Q_2^3 Q_3^5 Q_4^5},
\nonumber \\
&&\Omega_{2,1}^{PS}=-\frac{\sqrt{3} \left(\alpha _3+\alpha _4-1\right) \left(4 Q_3^2-Q_4^2\right)}{2
   Q_2^3 Q_3^5 Q_4^5},
\nonumber \\
&&\Omega_{3,1}^{PS}=-\frac{\sqrt{3} \alpha _4 \left(4 Q_3^2+Q_4^2\right)}{2 Q_2^3 Q_3^5 Q_4^5},
\nonumber \\
&&\Omega_{1,2}^{PS}=-\frac{\sqrt{\frac{3}{2}} \left(\alpha _2+\alpha _3-2 \alpha _4\right) \left(4
   Q_3^2-Q_4^2\right)}{4 Q_2^3 Q_3^5 Q_4^5},
\nonumber \\
&&\Omega_{2,2}^{PS}=\frac{\sqrt{\frac{3}{2}} \left(\alpha _3+\alpha _4-1\right) \left(2
   Q_3^2+Q_4^2\right)}{Q_2^3 Q_3^5 Q_4^5},
\nonumber \\
&&\Omega_{3,2}^{PS}=\frac{\sqrt{\frac{3}{2}} \alpha _4 \left(2 Q_3^2-Q_4^2\right)}{Q_2^3 Q_3^5
   Q_4^5},
\nonumber \\
&&\Omega_{1,3}^{PS}=-\frac{3 \left(\alpha _2-\alpha _3\right) \left(4 Q_3^2-Q_4^2\right)}{4 \sqrt{2}
   Q_2^3 Q_3^5 Q_4^5},
\nonumber \\
&&\Omega_{2,3}^{PS}=-\frac{3 \left(\alpha _3+\alpha _4-1\right)}{2 \sqrt{2} Q_2^3 Q_3^5 Q_4^3},
\nonumber \\
&&\Omega_{3,3}^{PS}=\frac{3 \alpha _4}{2 \sqrt{2} Q_2^3 Q_3^5 Q_4^3},
\end{eqnarray}
where
\begin{eqnarray}
&& Q_2^2=\frac{R_M^2}{2}+R_{uuds\bar{s}}^2,
\nonumber \\
&& Q_3^2=\frac{1}{4} \left(2 R_M^2+3 \left(R_B^2+R_{uuds\bar{s}}^2\right)\right),
\qquad\qquad\qquad\qquad\qquad\qquad\qquad
\nonumber \\
&& Q_4^2=R_B^2+R_{uuds\bar{s}}^2,
\nonumber \\
&& \alpha_2=0,
\nonumber \\
&& \alpha_3=-\frac{-R_B^2-R_{uuds\bar{s}}^ 2}{2 R_M^2+3 R_B^2+3
   R_{uuds\bar{s}}^2},
\nonumber \\
&& \alpha_4=-\frac{-R_M^2-R_B^2-R_{uuds\bar{s}}^2}{2 R_M^2+3 R_B^2+3
   R_{uuds\bar{s}}^2},
\nonumber \\
&&  \beta_2=1/2~,
\nonumber \\
&& \beta_3=-\frac{R_M^2+3
   R_B^2+R_{uuds\bar{s}}^2}{2
   R_M^2+3 R_B^2+3
   R_{uuds\bar{s}}^2}
   ,
\nonumber \\
&& \beta_4=-\frac{R_M^2+2
   R_{uuds\bar{s}}^2}{2
   \left(2 R_M^2+3 R_B^2+3
   R_{uuds\bar{s}}^2\right)}.
\end{eqnarray}

In this work we choose the radial parameters for the baryons and mesons as $R_B=R_{uuds\bar{s}}=3.1 ~GeV^{-1}$, $R_M=4.1~ GeV^{-1}$
\cite{Gutsche:1997gy}.

The initial state $|i\rangle $ and final state $|f\rangle$ can be written as
\begin{eqnarray}\label{state-i}
|i\rangle=|\{ \chi_{\frac{1}{2},m_{5q}}(uuds\bar{s })\otimes
\chi_{\frac{1}{2},m_{\bar{p }}}
(\bar{u}\bar{u}\bar{d})\}_{S,S_z}\otimes (L,M)\rangle_{J,J_z},
\end{eqnarray}
\begin{eqnarray}\label{state-f}
|f\rangle=| \{\chi_{1,m_\alpha}(\phi)\otimes
\chi_{j_m,m_{3',4'}}(X)\}_{j,m_\epsilon }\otimes
(\ell_f,m_f)\rangle_{J,J_z},
\end{eqnarray}
where $\chi_{\frac{1}{2},m_{5q}}(uuds\bar{s })$ is the spin-flavor-color part of the $5q$
component, $L $ and $l_f$ are respectively the initial and final orbital angular momenta, $J $ is the total angular momentum, and $I$ is the isospin.
The matrix element $ \langle f| O_{A_1}|i \rangle$ can be evaluated by using the two-body matrix elements for spin, flavor and color, corresponding to the $^3P_0$ quark model,
\begin{equation}\label{3p0-spin}
\langle 0 |\sigma^{ij }_\upsilon | \chi^{{J_{ij }}}_{m_{ij }}(ij) \rangle=\delta_{J_{ij },1}\delta_{m_{ij },-\upsilon}(-1)^\upsilon\sqrt{2},
\end{equation}
\begin{equation}\label{3p0-flavor}
\langle 0 |1^{ij }_F | \chi^{{T_{ij }}}_{t_{ij }}(ij) \rangle=\delta_{T_{ij },0}\delta_{t_{ij },0}\sqrt{2},
\end{equation}
and
\begin{equation}\label{3p0-color}
\langle0|1^{ij}_{C}|q_{\alpha}^i\bar{q}_\beta^j\rangle=\delta_{\alpha\beta},
\end{equation}
where $\alpha$ and $\beta$ are the color indices.
\begin{table}\label{BRtable[s to p]}
\caption{Branching ratio $BR(\times 10^{4})$ for the transition
$p\bar{p}\rightarrow \phi X$ ($X =\pi^0,\eta,\rho^0,\omega$) in
$p\bar{p}$ $s$-wave states annihilation at rest with the initial state is denoted by $^{2I+1,2S+1} L_J$. The results indicated by $\star$, have been normalized to the experimental values.}
\begin{center}
\begin{tabular}{c c c c c c c c}
\hline
\hline
&&&\\
 & \small{$^{11}S_0$$\rightarrow\omega\phi$} & \small{$^{33}S_1$$\rightarrow\pi^0\phi$}
& \small{$^{31}S_0$$\rightarrow\rho^0\phi$}&\small{$^{13}S_1$$\rightarrow\eta\phi$} \\
&&&\\
\hline
&&&\\
BR$^{\rm exp}$ & 6.3$\pm$2.3  & 5.5 $\pm$ 0.7& 3.4 $\pm$ 1.0 & 0.9 $\pm$ 0.3& \\
&&&\\

$[4][22][22]$& 6.3$\star$& 5.4 & 3.8 & 1.4 - 1.8 & \\
&&&\\
$[4][31][31]$& 6.3$\star$& 5.4 & 3.8 & 1.4 - 1.8 & \\
&&&\\
$[31][211][22]$& 6.3$\star$& 4.3 & 3.8 & 1.9 - 2.5 & \\
&&&\\
$[31][211][31]$& 6.3$\star$& 3.8 & 2.7 & 1.2 - 1.5 & \\
&&&\\
$[31][22][31]$& 6.3$\star$& 5.9 & 4.9 & 1.0 - 1.4 & \\
&&&\\
$[31][31][22]$& 6.3$\star$& 7.3 & 3.9 & 0.90 - 1.0 & \\
&&&\\
$[22][211][31]$& 6.3$\star$& 181.2 & 97.5 & 44.0 - 57.6 & \\
&&&\\
$[31][31][31]$& 6.3$\star$& 10.0 & 6.3 & 2.7 - 3.6 & \\
&&&\\
$[22][22][22]$& 6.3$\star$& 0.85 & 3.4 & 5.3 - 6.9 & \\
&&&\\
$[211][211][22]$& 6.3$\star$& 0.85 & 3.5 & 5.3 - 6.7 & \\
&&&\\
$[211][211][31]$& 6.3$\star$& 7.7 & 4.0 & 3.2 - 4.2 & \\
&&&\\
$[22][31][31]$& 6.3$\star$& 4.5 & 2.3 & 2.2 - 2.9 & \\
&&&\\
$[211][22][31]$& 6.3$\star$& 181.2 & 97.5 & 44.0 - 57.6 & \\
&&&\\
$[211][31][22]$& 6.3$\star$& 0.85 & 3.5 & 5.3 - 6.9 & \\
&&&\\
$[211][31][31]$& 6.3$\star$& 0.24 & 0.17 & 0.090 - 0.11 & \\
&&&\\

\hline
\hline
\end{tabular}
\end{center}
\end{table}

\begin{table}\label{BRtable[p to s]}
\caption{Branching ratio $BR(\times 10^{4})$ for the transition
$p\bar{p}\rightarrow \phi X$ ($X =\pi^0,\eta,\rho^0,\omega$) in
$p\bar{p}$ $p$-wave states annihilation at rest.}
\begin{center}
\begin{tabular}{c c c c c c c c}
\hline
\hline
&&&\\
 & \small{$^{33}P_{0,1,2}$$\rightarrow\rho^0\phi$} & \small{$^{31}P_1$$\rightarrow\pi^0\phi$}
& \small{$^{13}P_{0,1,2}$$\rightarrow\omega\phi$}&\small{$^{11}P_1$$\rightarrow\eta\phi$} \\
&&&\\
\hline
&&&\\
BR$^{\rm exp}$ & 3.7$\pm$0.9  & 0 $+$ 0.3& 2.9 $\pm$ 1.4 & 0.4 $\pm$ 0.2& \\
&&&\\

$[4][22][22]$& 3.7$\star$& 0.31 & 1.6 & 0.10 - 0.14 & \\
&&&\\
$[4][31][31]$& 3.7$\star$& 1.1 & 5.3 & 0.36 - 0.47 & \\
&&&\\
$[31][211][22]$& 3.7$\star$& 0.48 & 1.3 & 0.17 - 0.22 & \\
&&&\\
$[31][211][31]$& 3.7$\star$& 1.4 & 6.1 & 0.55 - 0.71 & \\
&&&\\
$[31][22][31]$& 3.7$\star$& 0.65 & 5.1 & 0.18 - 0.23 & \\
&&&\\
$[31][31][22]$& 3.7$\star$& 0.22 & 2.5 & 0.067 - 0.087 & \\
&&&\\
$[22][211][31]$& 3.7$\star$& 0.012 & 0.23 & 1.2-1.5 ($\times10^{-5}$) & \\
&&&\\
$[31][31][31]$& 3.7$\star$& 0.76 & 2.4 & 0.26 - 0.34 & \\
&&&\\
$[22][22][22]$& 3.7$\star$& 0.0029 & 13 & 0.0061 - 0.0080 & \\
&&&\\
$[211][211][22]$& 3.7$\star$& 0.0029 & 13 & 0.0061 - 0.0080 &  \\
&&&\\
$[211][211][31]$& 3.7$\star$& 0.0029 & 3.6 & 0.23 - 0.30 & \\
&&&\\
$[22][31][31]$& 3.7$\star$& $5.1\times10^{-4}$ & 2.2 & 0.16 - 0.20 & \\
&&&\\
$[211][22][31]$& 3.7$\star$& 0.012 & 0.23 & 1.2-1.5 ($\times10^{-5}$) & \\
&&&\\
$[211][31][22]$& 3.7$\star$& 0.0029 & 13 & 0.0061 - 0.0080  & \\
&&&\\
$[211][31][31]$& 3.7$\star$& 7.4$\times10^{-4}$ & 0.63 & 0.0062 - 0.0081 & \\
&&&\\
\hline
\hline
\end{tabular}
\end{center}
\end{table}

Since we consider $p\bar{p}$ annihilations at rest, the proton-antiproton wave function is strongly correlated due to the $N \bar N$ interaction \cite{Carbonell:1989cs,Kang:2013uia}. Therefore the initial state interaction for the atomic state of the $p\bar{p}$ system has to be involved \cite{Kercek:1999sc}, resulting in the transition amplitude \begin{equation}\label{atomic state}
T_{f,i}(\vec{q})=\int d^3k ~ T^{SP(PS)}_{fi}(\vec{q},\vec{k}) \phi ^I_{LSJ}(\vec{k} ),
\end{equation}
where $\phi ^I_{LSJ}(\vec{k} ) $ is the protonium wave function in the momentum space for fixed isospin $I $. With the transition amplitude, the partial decay width for the transition of $p\bar{p} $ atomic states to two-meson final states $\phi X $ can be calculated by
\begin{equation}
\Gamma_{p\bar{p}\rightarrow \phi X}=\frac{1}{2E}\int\frac{d^3p_\phi}{2E_\phi}\frac{d^3p_X}{2E_X} ~\delta ^{(3)}(\vec{p}_\phi +\vec{p}_X)\delta(E-E_\phi-E_X )|T_{f,i}(\vec{q})|^2\; ,
\end{equation}
where $E$ is the total energy ($E=1.876$ GeV) and $E_{\phi ,X}= \sqrt{m^2_{\phi ,X}+\vec{p}^2_{\phi ,X}} $ is the energy of the outgoing mesons $\phi$ and $X$ with mass $m_{\phi ,X}$ and momentum $\vec{p}_{\phi ,X}$. With the obtained transition amplitude given by Eqs (\ref{T-1}) and (\ref{atomic state}), the partial decay width for the transition from the $p\bar{p} $ atomic state $|i\rangle=|ILSJ\rangle$ can be written as
\begin{equation}\label{decay width}
\Gamma_{p\bar{p}\rightarrow \phi X}=|AB|^2\lambda_{A_1}^2f(\phi ,X)\langle f | O_{A_1}|i \rangle ^2 \gamma_i(I) .
\end{equation}
Here, the function $f(\phi ,X)$ is the kinematical phase-space factor depending on the
relative momentum and the masses of $\phi X$ system while $\gamma_i(I)$ is the factor depending on the initial-state interaction. Thus, the branching ratio $BR$ of the annihilation reactions at rest $p\bar{p} $ $ \rightarrow \phi X$($X=\pi^0,\eta,\rho^0,\omega$) can be expressed as
\begin{equation}\label{BR}
BR_i(\phi,X)=\frac{ (2J+1)\Gamma_{p\bar{p}\rightarrow \phi X}}{\Gamma_{tot}(i)},
\end{equation}
where $(2J+1)$ are the statistical weights corresponding to the initial values of the total angular momentum $J$. The fraction $\Gamma_{tot}(i) $ denotes the total annihilation width of the $p\bar{p} $ atomic state with fixed principal quantum number~\cite{Dover:1991mu}.
Nevertheless, the model dependence due to the harmonic oscillator approximation may be reduced by applying a simplified phenomenological approach for $N \bar N$ annihilation \cite{Kercek:1999sc,Gutsche:1998fc}. In stead of the obtained kinematical phase-space factor $f(\phi ,X)$, we use the phenomenological form $ f(\phi,X)=q\cdot {\rm exp}\{-(1.2)~GeV^{-1}\,(s-s_{\phi X})^{1/2}\}$ with $s_{\phi X}=(m_{\phi}+m_{X})^{1/2}$ and
$\sqrt{s}=(m_{\phi}^2+q^2)^{1/2}+(m_{X}^2+q^2)^{1/2} $ \cite{Vandermeulen:1988hh}. For the functions $\gamma_i(I) $, depending on the initial-state interaction, are related to the probability for a protonium state to have isospin $I$ and spin $J $. We adopt the probability $\gamma_i(I) $ and the total decay width $\Gamma_{tot}(J) $ obtained in an optical potential calculation \cite{Dover:1991mu,Carbonell:1989cs}.

The obtained theoretical results for branching ratios, of Eq.(\ref{BR}) for each $uuds \bar{s}$ configurations, of the $S$ to $P$ ($L=0,$ $\ell_f=1$) and $P$ to $S$  ($L=1,$ $\ell_f=0$) transitions are compared with the experimental data (BR$^{\rm exp}$) in Table 3 and 4, respectively. To eliminate the factor $|AB|^2\lambda_{A_1}^2$ which is unknown priori, the model predictions one entry (as indicated by $\star $) have been normalized to the experimental number. Therefore, the obtained branching ratios can not be for estimating the pentaquark content (the coefficient $B$) in the nucleon.  For the transition $p\bar p\rightarrow \phi \eta $, the physical $\eta $ meson is produced by its nonstrange component $\eta_{ud} $ with $\eta = \eta_{ud}(\sqrt{1/3}\cos\theta - \sqrt{2/3}\sin\theta) $ with the pseudoscalar mixing angle $\theta $ variate from $-10.7^o $ to $-20^o$. As shown in Table 3 and 4, the model predictions with the flavor-spin mixed symmetry $[4]_{FS}$ are in good agreement with the experimental data. Especially, excellent agreement is found in the configuration with flavor-spin symmetry $[4]_{FS}[22]_F[22]_s$ which gave the branching ratios consistent with the experimental data for both the $S$ to $P$ and $P$ to $S$ transitions.

\section{Summary} \label{sec:3}

 We have estimated the branching ratios of the annihilation reactions at rest $p\bar{p} \rightarrow \phi X$ ($X=\pi^0,\eta,\rho^0,\omega$) from atomic $p\bar{p}$ $S$- and $P$-wave states in the effective quark line diagrams incorporating the $^3P_0$ model. The proton wave function are assumed including the intrinsic strangeness in the form of $qqqs\bar{s}$ components. Considered in the work are 15 $qqqs\bar{s}$ configurations, where the $\bar{s}$ is in the ground state but the subsystem $uuds$ is in the $P$-state since these configurations lead to negative strangeness spin $\sigma_s $ and positive magnetic moment $\mu_s $.

 It is found in Table 3 that the theoretical results with the flavor-spin symmetries $[4]_{FS}[22]_F[22]_s$, $[4]_{FS}[31]_F[31]_s$, $[31]_{FS}[211]_F[31]_s$, $[31]_{FS}[22]_F[31]_s$ and $[31]_{FS}[31]_F[22]_s$ for the pentaquark components are consistent with the experimental data for $p\bar p$ annihilation in the $S$-wave.
 Table 4 shows that for $p\bar p$ annihilation in the $P$-wave the pentaquark configurations with the flavor-spin symmetries $[4]_{FS}[22]_F[22]_s$ and $[31]_{FS}[211]_F[22]_s$ lead to theoretical predictions consistent with the experimental data. Therefore, one may conclude that the best agreement of theoretical results with the experimental data is found in the pentaquark configuration with the flavor-spin symmetry $[4]_{FS}[22]_F[22]_s$.
\\

{\bf Acknowledgements} {\\ This work was supported by Suranaree University of Technology (SUT) under Grant No. SUT1-105-57-12-27 and SUT-CHE-NRU project (NV. 4/2558). SS acknowledges support from Faculty of Science, Burapha University.}


\end{document}